\documentclass[aps,prl,twocolumn,showpacs,groupedaddress]{revtex4-1}

\usepackage{graphicx}
\usepackage{dcolumn}
\usepackage{bm}

\begin{document}

\title{Optical control of coherent interactions between quantum dot electron spins}

\author{S.\,Spatzek$^1$}
\author{A.\,Greilich$^{1}$} \email[Electronic
address: ]{alex.greilich@udo.edu}
\author{Sophia\,E.\,Economou$^2$}
\author{S.\,Varwig$^1$}
\author{A.\,Schwan$^1$}
\author{D.\,R.\,Yakovlev$^{1,3}$}
\author{D.\,Reuter$^4$}
\author{A.\,D.\,Wieck$^4$}
\author{T.\,L.\,Reinecke$^2$}
\author{M.\,Bayer$^1$}

\affiliation{$^1$Experimentelle Physik 2, Technische Universit\"{a}t Dortmund, D-44221 Dortmund, Germany}
\affiliation{$^2$Naval Research Laboratory, Washington, D.C. 20375, USA}
\affiliation{$^3$A.~F. Ioffe Physico-Technical Institute, RAS, St. Petersburg, 194021 Russia}
\affiliation{$^4$Angewandte Festk\"{o}rperphysik, Ruhr-Universit\"{a}t Bochum, D-44780 Bochum, Germany}

\date{\today}

\begin{abstract}
Coherent interactions between spins in quantum dots are a key
requirement for quantum gates. We have performed pump-probe
experiments in which pulsed lasers emitting at different photon
energies manipulate two distinct subsets of electron spins within an
inhomogeneous InGaAs quantum dot ensemble. The spin dynamics are
monitored through their precession about an external magnetic field.
These measurements demonstrate spin precession phase shifts and
modulations of the magnitude of one subset of oriented spins after
optical orientation of the second subset. The observations are
consistent with results from a model using a Heisenberg-like
interaction with $\mu$eV-strength.
\end{abstract}


\maketitle

Considerable progress has been made recently in establishing optical
control of spins confined in semiconductor quantum dots (QDs), a
system of interest for quantum bits (qubits) in implementations of
quantum information~\cite{Bur00}. Single spin decoherence times on
the order of microseconds have been demonstrated~\cite{Gre06s}, and
methods for spin initialization and readout have been
developed~\cite{Ata06,Xu07}. Recently, progress in demonstrating
optical rotations of single spins has been
made~\cite{Wu07,Ber08,Press08,Gr09}. To be useful in quantum
information, spin manipulation times must be orders of magnitude
faster than decoherence times~\cite{Bur00}, which is possible using
fast optical methods. Interactions between spins in QD systems can
provide the mechanism for coherent control in quantum logic but can
also complicate their coherent dynamics. The case of coupling
between spins in QD molecules has been well
studied~\cite{Sti06,Kren05,Ortn05,Kim10}, but long ranged
interactions are not yet understood.

An ensemble of QDs has the advantage of having strong optical
coupling, but ensemble approaches typically have been hampered by
inhomogeneities in their properties, particularly spin splittings,
which lead to fast spin dephasing. In previous work we have
demonstrated nuclear assisted optical techniques for removing some
of the effects of these inhomogeneities~\cite{Gre06s,Gr07}. In these
techniques, periodic pulse trains orient spins normal to an external
magnetic field, and particular subsets of spins precess in phase
with the pulse trains. At rather low magnetic fields, around $B$ =
1\,T, a spin ensemble can be put into a state in which only few spin
precession modes contribute~\cite{Gr09b}. This is the system that we
study here.

In the present work two subsets of spins are selected by spectrally
narrow, circularly polarized laser pulse trains of different photon
energies. The subsets are oriented by the two laser pulses (pump 1
and pump 2), and precess around a perpendicular magnetic field. The
relative phase of the two precessions is controlled by the time
difference between the two pulses. We find that after the second
pump pulse, the precession associated with the first spin subset
acquires a phase shift that depends on the relative orientation of
the spins. It emerges smoothly in time after pump 2. In addition,
the precession amplitude shows modulations and decreases with time.
The major experimental features are consistent with a
Heisenberg-like interaction between the spins in the ensembles with
strength on the order of $\mu$eV.

\begin{figure}
\includegraphics[width=\columnwidth]{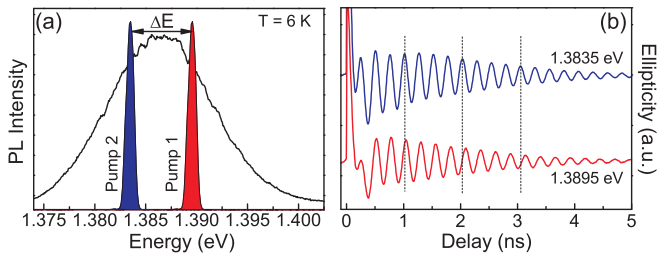}
\caption{\label{fig:PL}(color online) (a) Photoluminescence spectrum of the (In,Ga)As QD sample. Two shaded areas
give lineshapes of the picosecond laser pulses at energies 1.3835\,eV and 1.3895\,eV used to initialize
two subsets of spins. (b) Ellipticity traces of the two different spin subsets. At early delays $<$500\,ps, the signals show some weak exciton interference from neutral QDs.}
\end{figure}

The experiments were performed on an ensemble of self-assembled
(In,Ga)As/GaAs QDs grown by molecular beam epitaxy such that each QD
contains on average a single electron. The sample contained 20
layers of dots with 60\,nm separation between adjacent layers and a
sheet dot density of ~10$^{10}$ cm$^{-2}$. The experiments were
performed at $T$ = 6\,K in a magnetic field of 1\,T. The spin
dynamics were investigated by time-resolved ellipticity, which
measures the spin projection along the optical axis (the z
direction), which coincides with the QD growth direction. The sample
was excited by two phase-synchronized trains of pump laser pulses
with a time jitter well below 1\,ps. The pump lasers were tuned to
different energies in the inhomogeneously broadened QD
photoluminescence, as sketched in Fig.~\ref{fig:PL}a. The laser
pulses were emitted at a frequency of 75.75\,MHz, and had durations
of 2\,ps corresponding to 1\,meV spectral width. The circular
polarizations of the two lasers were adjusted independently, as was
the delay between them. For ellipticity studies a weak probe was
split from one of the pumps, and after being polarized linearly it
was sent through the sample. The change of probe polarization
ellipticity was recorded by a balanced detection
scheme~\cite{Gre06s}.

The circularly polarized optical pulse of intensity $\pi$ excites
the QD spin to a trion state leaving the other spin to precess
around a perpendicular magnetic field applied along the $x$
direction~\cite{Gr06}. Figure~\ref{fig:PL}b gives ellipticity
results with a single pump laser exciting the QD ensemble that is
probed at the same energy. In the upper trace the pump and probe
photon energy were on the low energy side of the photoluminescence
band, and in the lower trace they were shifted to the high energy
side by $\Delta E \sim$6\,meV. In both cases the pump laser creates
spin coherence at time zero after which the electron spin precesses.
We estimate that there are about 10$^6$ spins in each subset
corresponding to an average separation exceeding 90\,nm between the
spins. Note that the precession frequencies are different from one
another due to the difference in their electron $g$-factors.

In the two-pump laser experiments about the same spacing as in Fig.~\ref{fig:PL}a was used for the two pump
energies, so that the pulses had no spectral overlap. The lasers therefore orient the spins in distinct subsets of
QDs. The pulses were sufficiently detuned so that no spin rotation of one spin subset by the laser
exciting the other subset could be resolved~\cite{Gr09,Econ07}. The signature of such a rotation would be an
instantaneous phase shift at the time of laser pulse.

\begin{figure}
\includegraphics[height=10cm]{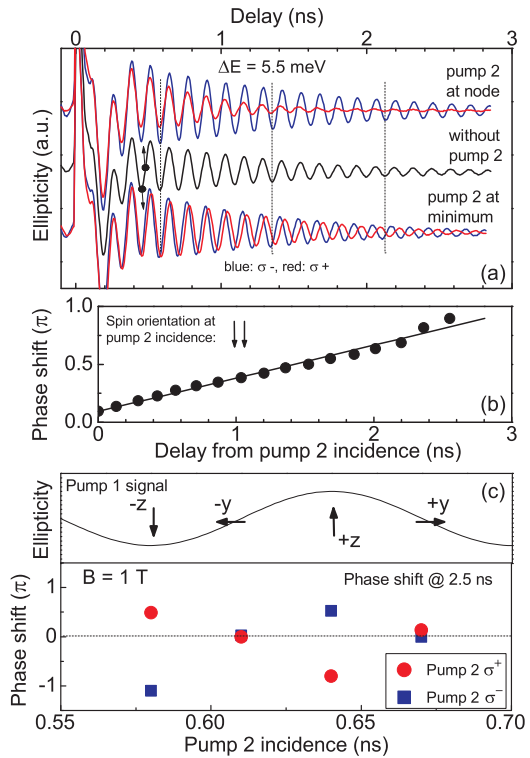}
\caption{\label{fig:Ell}(color online) (a) Ellipticity as function of time delay for two pumps. (b) Phase shift evolution in time after pump 2 for collinear spin orientations.
(c) Phase shift as function of incidence times of pump 2. Black curve is reference trace, and arrows indicate orientations of spin subset 1 when pump 2 is applied.
Phase shifts are measured at 2.5\,ns probe delay.}
\end{figure}

Figure~\ref{fig:Ell}a gives results when two circularly polarized
pulses are applied with a fixed time difference between them for
each trace. The probe energy used to measure the spin coherence was
the same as that of pump 1. Thus the effect of the spins driven by
pump 2 on those driven by pump 1 is monitored. The black curve is a
reference trace with only pump 1 on. The incidence times of pump 2
are given by the dots on the reference trace. We focus first on the
2\,ns right after pump 2.

For the bottom pair of traces in Fig.~\ref{fig:Ell}a, pump 2 is applied when the reference trace is at a minimum
so that spin subset 1 is pointing in the $-z$ direction. Red (grey) and blue (dark grey) traces are for the two circular
polarizations of pump 2 creating spins that point along the $+z$ or the $-z$ direction, respectively. Phase shifts
with respect to the reference emerge after pump 2 and have opposite signs for the two pump 2 polarizations. The
slow, nearly linear emergence of phase shifts with increasing probe delay time after pump 2 is shown in the
Fig.~\ref{fig:Ell}b. This slow time dependence excludes its resulting from rotation of subset 1 by laser 2.  The
top pair of traces corresponds to pump 2 being applied when the ellipticity signal is zero, i.e., when spin subset
1 points along the $-y$ direction. In this case the phase shifts are small.

\begin{figure}
\includegraphics[width=\columnwidth]{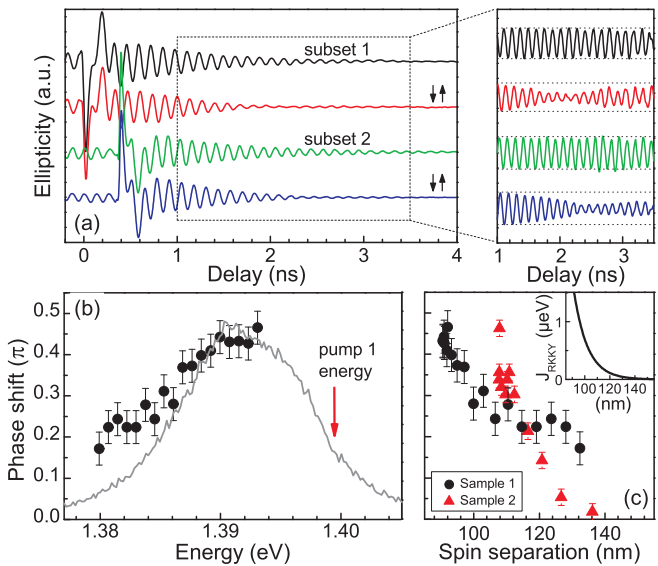}
\caption{\label{fig:Pol}(color online) (a) Left panel gives
ellipticity for single pump-probe measurements of spin subset 1 at
1.39\,eV (black) and spin subset 2 at 1.385\,eV (green, 3rd from
top). Red (2nd) and blue (4th) curves are for two pumps with probe
at the energy of pump 1 or 2, respectively. The right panel gives
zoom-in of time dependences with exponential decay component removed
(see text). (b) Phase shift as a function of detuning between pumps
at one sample position. (c) Phase shift as a function of spin
separation for two measured samples. (Inset) interaction strength
$J_{RKKY}$ as a function of spin separation.}
\end{figure}

The dependences of the phase shifts on the polarizations of pumps 1
and 2 are given in Fig.~\ref{fig:Ell}c. Pump 1 had $\sigma^+$
polarization, and pump 2 had $\sigma^+$ or $\sigma^-$ polarizations.
The incidence time of pump 2 was varied to provide different
orientations of spin subset 1, which are indicated by the black
arrows on the top. The phase shifts are essentially zero when spin
subsets 1 and 2 are perpendicular at pump 2. They are large when
subset 1 is along $+z$ or $-z$, and they are of opposite sign for
$\sigma^+$ and $\sigma^-$ polarizations of pump 2.

Additional interesting features appear when the signal is monitored
over longer delays up to 4\,ns for the cases of large phase shifts.
These results are shown in Fig.~\ref{fig:Pol}a. The black (top) and
green (3rd from the top) traces in the left panel give the
ellipticity after excitation by a single pump laser so that only
spin subset 1 or subset 2 is oriented. The separation of 390\,ps
between the two pumps is the same as in the two-pump experiments
described below. In the single pump experiments we observe a decay
of the envelope of the $z$ spin component with increasing delay. We
associate this decay with dephasing due to inhomogeneous spin
precession. This dephasing is weak because at 1\,T magnetic field
each spin subset precesses with a number of modes close to
one~\cite{Gr07}. The ellipticity envelope after trion decay can be
fitted accurately by an exponential with a dephasing time $T_2^*$ =
0.8\,ns.

The red (2nd from top) and blue (4th) traces in Fig.~\ref{fig:Pol}a give the ellipticities when both pump 1 and pump 2 are applied,
with the probe on spin subset 1 for the red (2nd) trace and on spin subset 2 for the blue (4th) trace. The initial directions
of spin subsets 1 and 2 are given by the arrows. In each case we see small, but clear modulations of the signal
near 2.5 - 3\,ns. These features are clearer when we remove the exponential decay due to dephasing~\cite{Gr06}
using the time $T_2^*$ from the single pump measurements. The results are shown to the right portion of
Fig.~\ref{fig:Pol}a. For the black and green (3rd) curves with only one pump, we see harmonic oscillations without
modulation. There are two distinct features from the red (2nd) and blue (4th) traces with two pumps: a modulation of the
magnitudes of the envelopes and a decay of the envelope of ellipticity compared to the one pump cases.

\begin{figure}
\includegraphics[width=\columnwidth]{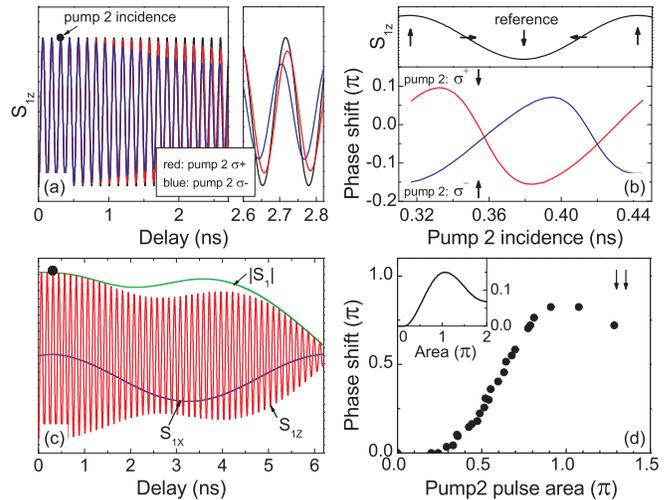}
\caption{\label{fig:theory}(color online) Calculations for: (a) Spin
polarization along the optical axis z as function of delay time from
pump 1. Right panel shows (zoom in) extended time period and phase
shifts. (b) Phase shifts at probe delay time of 2.5\,ns for
$\sigma^+$ and $\sigma^-$ polarizations of pump 2; incidence times
such that spin 1 is oriented as shown by arrows in the top panel.
(c) Expectation values of $S_{1z}$, $S_{1x}$ and $|S_1|$. (d)
Measured phase shift as a function of pump 2 power. (Inset)
calculated power dependence of the phase shifts.}
\end{figure}

The power dependence of the phase shifts is shown in
Fig.~\ref{fig:theory}d, where the phase shift increases with power
up to pulse area of $\pi$ and decreases thereafter.

In order to understand these results, we consider a spin system with
optical pulses and with interactions between the spins. For
simplicity we consider a model of two spins interacting with a
Heisenberg form $J\mathbf{S}_1 \cdot \mathbf{S}_2$ where the spins
are subject to separate periodic optical pulse trains of different
energies. Here $J$ is the interaction strength, and $\mathbf{S}_1$,
$\mathbf{S}_2$ are the spin operators.

We solve for the steady state dynamics of the system, which is done
by constructing the evolution operator for the combined system,
obtaining the corresponding density matrix and propagating it
forward in time to the joint steady state. The expectation value of
spin 1 as a function of time is obtained by tracing out spin 2 from
the density matrix. Dephasing from the environment is not included
here, and as a result there is no loss of amplitude in time from
sources outside of the spin system. We have considered the effects
of other unpolarized spins using a simple model and find that they
do not affect the qualitative features of the response~\cite{SOM}.

Results from these calculations are given in Fig.~\ref{fig:theory}.
Figure~\ref{fig:theory}a gives the time dependence of expectation
value $S_{1z}$. The black reference trace is for only pump 1
applied. The dot indicates time of pump 2, when reference spin is in
$+z$ direction. Red (grey) and blue (dark grey) traces are for
$\sigma^+$ and $\sigma^-$ polarizations for pump 2. The extended
panel in Fig.~\ref{fig:theory}a shows the two phase shifts emerging
smoothly and approximately linearly in time after pump 2.
Fig.~\ref{fig:theory}b gives the phase shifts for varying times of
application of pump 2 for the two pump 2 polarizations. We see that
the phase shifts for the two polarizations of pump 2 are opposite in
sign. These phase shifts are absent without the interaction $J$.
Asymmetries in the features arise from the two $g$-factors being
unequal.

Calculated results for a wider range of delay times are shown in
Fig.~\ref{fig:theory}c. The curves give $S_{1z}$, $S_{1x}$ and
$|S_1|$ of spin 1 after spin 2 is traced out of the density matrix.
$S_{1z}$ oscillates around the magnetic field with the Larmor
frequency. In the absence of interactions between the spins, the
envelopes of $S_{1z}$, and $|S_1|$ would be constant in time, and
$S_{1x}$ would be zero. With interactions, the envelope of $S_{1z}$
decreases in time and becomes modulated. In addition, the overall
magnitude $|S_1|$ decreases in time. These features result from
coupled dynamics of the two spins in the presence of the
interaction. Inset to Fig.~\ref{fig:theory}d shows the calculated
phase shift as a function of pump 2 power. This behavior results
from an oscillation of the spin polarization of subset 2 excited by
pump 2, which subsequently interacts with spin subset 1.

We find that a value of $J \sim 1 \mu$eV gives features qualitatively similar to those in experiment in
Fig.~\ref{fig:Ell}. We have also tried other forms of interactions between spins, including an Ising form. These
forms give a number of results similar to those from the Heisenberg interaction but are in less good overall
accord with experiment.

We see that the key features from experiment are consistent with the
results of this model of interacting spins. The ellipticity in the
experiments corresponds to the spin magnitude in the model. (i) In
both cases the phase shifts emerge smoothly in time after the second
optical pulse. In the model this behavior arises from the coupled
dynamics of the two interacting spins, and it would not be present
without the interaction. (ii) The dependence of the phase shifts on
the polarizations of the two lasers is similar. In both cases for a
fixed polarization of pump 1, the phase shifts are opposite in sign
for $\sigma^+$ and $\sigma^-$ polarizations of the pump 2. In both
cases the phase shifts are large when the spins are either parallel
or antiparallel and small when they are perpendicular at the second
pulse. (iii) The dependence of the phase shifts on the power of pump
2 in Fig.~\ref{fig:theory}d is similar in experiment and model. In
both cases the phase shift increases from low power, reaches a
maximum near a pulse of $\pi$ and decreases after that. The fact
that phase shift of subset 1 follows the degree of spin polarization
of subset 2 is associated with spin interactions. (iv) The
modulations in magnitude of the ellipticity in experiment correspond
to the modulations of the spin magnitude in the model. These
features result from the coupled dynamics of the spins in the
presence of interactions.

From this list of similar features in experiment and theory we
conclude that the experimental results give convincing evidence for
existence of interactions between the spins in these QD arrays.

The presence of interactions between spins is given added support
from results for the dependence of the phase shifts on the
separation between dots. The phase shifts measured at a fixed
position on the sample as functions of the detuning between the two
pumps are shown in Fig.~\ref{fig:Pol}b with the corresponding
photoluminescence spectrum. The arrow gives the position of pump 1,
and the black dots the positions of pump 2. The photoluminescence
intensities at pump 2 provide a measure of the number of dots
excited at the several energies.

The average separation between the spins is estimated from the
number of excited dots. To do so, we include explicitly the
separations of a given spin to spins within the layer and to those
in two adjacent layers~\cite{SOM}. We find that including more
distant spins does not affect the results significantly. The
fraction of the dots that overlap the laser spectrally is determined
by integrating the relevant regions of the photoluminescence
spectrum. We determine the ratio of uncharged dots to the singly
charged dots for each transition energy from the magnitudes of the
Faraday rotations before and immediately after pump pulse
application. The optically oriented electron spin density is
obtained from the optically excited dot density at each energy using
this ratio. Finally, the average separation between spins is
obtained from statistical averaging assuming that the dot
distribution in a layer is Poissonian~\cite{SOM}.

The phase shifts as functions of the average spin separation are
shown by the black symbols in Fig.~\ref{fig:Pol}c. The phase shifts
decrease for increasing spin separation, as expected for a
long-ranged interaction between spins. To support these results, an
additional sample was studied (sample 2), which has a dot density
four times higher than the first sample and a smaller interlayer
separation of $30$\,nm. Results from this sample are shown by the
red triangles in Fig.~\ref{fig:Pol}c. The somewhat larger spin
separation in that sample results from its larger spectral width and
higher probability of doubly charged dots in it.

The present understanding of spin dynamics of these inhomogeneous
arrays does not permit a definitive determination of the microscopic
interaction mechanism between spins. Nevertheless, among all of the
long-ranged spin interactions available~\cite{dip}, the optical RKKY
interaction discussed in Refs.~\cite{Pier02} and~\cite{Ram05} is the
only one that has an overall magnitude consistent with the value of
interaction $J \sim 1 \mu$eV obtained from experiment. To explore
this further, we plot in the inset of Fig.~\ref{fig:Pol}c the
dependence of the optical RKKY interaction, $J_{RKKY}$, between
spins calculated as described in Ref.~\cite{Ram05}. The resulting
dependence of $J_{RKKY}$ on spin separation is similar to the
dependence of the observed phase shift on average spin separation.
In addition, we note that the magnitude of $J_{RKKY}$ is in the
right $\mu$eV-range  for the parameters used~\cite{calc}.

In summary we have presented evidence for coherent interactions
between spins from pump-probe experiments on ensembles of InGaAs QDs
and from calculations. These interactions can play important roles
in coupling spins in quantum gates and in extended architectures for
quantum information

\begin{acknowledgments}
This work was supported by the BMBF project QuaHL-Rep, the Deutsche
Forschungsgemeinschaft, and the US Office of Naval Research.
\end{acknowledgments}


\begin{references}

\bibitem{Bur00}
G. Burkard, H. A. Engel, D. Loss, Prog.\ Phys. {\bf 48}, 965 (2000).

\bibitem{Gre06s}
A. Greilich {\it et al.}, Science {\bf 313}, 341 (2006).

\bibitem{Ata06}
M. Atat\"{u}re {\it et al.}, Science {\bf 312}, 551 (2006).

\bibitem{Xu07}
X. Xu {\it et al.}, Phys.\ Rev.\ Lett. {\bf 99}, 097401 (2007).

\bibitem{Wu07}
Y. Wu {\it et al.}, Phys.\ Rev.\ Lett. {\bf 99}, 097402 (2007).

\bibitem{Ber08}
J. Berezovsky {\it et al.}, Science {\bf 320}, 349 (2008).

\bibitem{Press08}
D. Press, T. Ladd, B. Zhang, and Y. Yamamoto, Nature {\bf 456}, 218 (2008).

\bibitem{Gr09}
A. Greilich {\it et al.}, Nat. Phys. {\bf 5}, 262 (2009).

\bibitem{Sti06}
E. A. Stinaff {\it et al.}, Science {\bf 311}, 636 (2006).

\bibitem{Kren05}
H. J. Krenner {\it et al.}, Phys.\ Rev.\ Lett. {\bf 94}, 057402 (2005).

\bibitem{Ortn05}
G. Ortner {\it et al.}, Phys.\ Rev.\ Lett. {\bf 94}, 157401 (2005).

\bibitem{Kim10}
D. Kim {\it et al.}, Nat. Phys. {\bf 7}, 223 (2011).

\bibitem{Gr07}
A. Greilich {\it et al.}, Science {\bf 317}, 1896 (2007).

\bibitem{Gr09b}
A. Greilich {\it et al.}, Phys.\ Rev.\ B {\bf 79}, 201305 (2009).

\bibitem{Gr06}
A. Greilich {\it et al.}, Phys.\ Rev.\ Lett. {\bf 96}, 227401
(2006).

\bibitem{SOM}
Supplemental material

\bibitem{Econ07}
S. E. Economou, and T. L. Reinecke, Phys.\ Rev.\ Lett. {\bf 99}, 217410 (2007).

\bibitem{Econ06}
S. E. Economou, L. J. Sham, Y. Wu, D. G. Steel, Phys.\ Rev.\ B {\bf 74}, 205415 (2006).

\bibitem{Pier02}
C. Piermarocchi, P. Chen, L.J. Sham, and D.G. Steel, Phys.\ Rev.\ Lett. {\bf 89}, 167402 (2002).

\bibitem{Ram05}
G. Ramon, Y. Lyanda-Geller, T.L. Reinecke, and L.J. Sham, Phys.\ Rev.\ B {\bf 71}, 121305(R) (2005).

\bibitem{dip}
We estimate that the classical dipolar interaction between spins for QD ensembles with average spin densities of
$\sim$10$^{10}$/cm$^2$ is much smaller, on the order of 10$^{-9}$ $\mu$eV, and interactions mediated by the nuclei
should have an upper bound of $\sim$10$^{-5}$ $\mu$eV due to slow nuclear spin diffusion between the dots

\bibitem{calc}
The calculations were made using a dot radius of 10\,nm, a lateral
confining potential of 150\,meV, a detuning of 0.5\,meV and optical
coupling of 0.5\,meV.

\end{references}
\end{document}